\documentclass[manuscript]{aastex}

\shorttitle{Extreme dust disks in Arp 220}
\shortauthors{Wilson et al.}

\begin{document}


\title{Extreme dust disks in Arp 220 as revealed by ALMA}


\author{C. D. Wilson\altaffilmark{1},
N. Rangwala\altaffilmark{2},
J. Glenn\altaffilmark{2},
P. R. Maloney\altaffilmark{2},
L. Spinoglio\altaffilmark{3},
\and
M. Pereira-Santaella\altaffilmark{3}}


\altaffiltext{1}{Department of Physics \& Astronomy, McMaster University,
    Hamilton, ON L8S 4M1 Canada}

\altaffiltext{2}{Center for Astrophysics and Space Astronomy, 389-UCB, University of Colorado, Boulder, CO 80303, USA}

\altaffiltext{3}{Istituto di Astrofisica e Planetologia Spaziali, INAF-IAPS, Via Fosso del Cavaliere 100, I-00133 Roma, Italy}


\begin{abstract}
We present new images of Arp 220 from 
the Atacama Large Millimeter/submillimeter Array
with the highest combination of frequency (691 GHz) and
resolution ($0.36 \times 0.20^{\prime\prime}$) ever obtained for this
prototypical ultraluminous infrared galaxy.
The western nucleus is
revealed to contain warm (200 K) dust that is optically thick ($\tau_{434\mu m} = 5.3$), while the eastern nucleus is cooler (80 K) and somewhat less opaque
($\tau_{434\mu m} = 1.7$).
 We derive full-width half-maximum diameters of $ 76 \times \le 70$
pc and $123 \times 79$ pc 
for the western and eastern nucleus, respectively. 
 The two nuclei combined 
account for ($83  ^{+65}_{-38}$ (calibration) $^{+0}_{-34}$ (systematic))\%
of the total infrared
luminosity of Arp 220.
The luminosity surface density of the western nucleus
($ \log(\sigma T^4) =  14.3\pm 0.2 ^{+0}_{-0.7}$ in units of L$_\odot$ kpc$^{-2}$)
appears sufficiently high to require the 
presence of an AGN or a ``hot starburst'', although the exact value depends
sensitively on the brightness distribution adopted for the source.
Although the role of any central AGN remains open, the inferred mean gas
column densities of $0.6-1.8 \times 10^{25}$ cm$^{-2}$ mean that any AGN
in Arp 220 
must be Compton-thick. 
\end{abstract}






\section{Introduction}

As the  closest example of an ultraluminous infrared galaxy
\citep{s84}, Arp 220
has been intensively studied at a wide range of wavelengths. 
Much of this work has been  focused on  determining the energy
sources powering its large infrared luminosity
\citep[$1.8\times 10^{12}$ L$_\odot$,][]{r11}. 
Infrared
spectroscopy has pointed to the dominance of a central starburst
\citep{g98,a07,c13}, with the contribution of a buried active galactic
nucleus (AGN) estimated to be 17\% \citep{n10}. Models of submillimeter
spectroscopy are not consistent with a dominant contribution from an
X-ray dominated region to the global CO emission \citep{r11}. However, 
high-resolution
X-ray data identified a source with a hard spectrum
coincident with the nuclear regions \citep{c02}.
Indications of HNC maser emission towards the western nucleus are also
suggestive of the presence of an AGN \citep{a09}.
In the radio, Arp 220  contains two compact nuclei \citep{bh95,rr05}, which
are also seen as peaks in the near-infrared extinction maps
\citep{s98,e11}.  High-resolution mid-infrared images show
that these two nuclei account for essentially all of the 24 $\mu$m
flux density \citep{so99}.

Arp 220 has been intensively studied at high resolution at millimeter 
wavelengths in both continuum and CO spectral lines
\citep{s91,s97, ds98,s99,de07,s08,m09,e11}. \citet{s99} conclude that most of
the far-infrared luminosity originates in the two nuclei, which each
contain a warm, dense gas disk $\sim 100$ pc in radius. 
\citet{de07} present a detailed analysis of the brighter western
nucleus and argue for the presence of an AGN based on the extremely
high surface brightness. They estimate a dust temperature of 90-170 K,
a radius of 35 pc, and an average gas density of $n_{H_2} = 9\times 10^4$
cm$^{-3}$. \citet{s08} also find dust temperatures of 90-160 K and
nuclear luminosities of $0.2-1 \times 10^{12}$ L$_\odot$, with the precise
values depending on the adopted source model. They conclude that
 the dominant energy source can be either an AGN or an extreme starburst.
\citet{m09} were also
able to distinguish the two nuclei at 435 $\mu$m.
All these studies find evidence for continuum optical depths
approaching unity at wavelengths of 0.87-1.3 mm. 

In this {\it Letter} we present new high frequency, high resolution
continuum imaging of Arp 220 
 ($D_L = 77.4$ Mpc, $D_A = 74.5$ Mpc) from the Atacama Large Millimeter/submillimeter
Array (ALMA). At a wavelength of 434 $\mu$m, the nuclear disks have
a significantly higher optical depth than the 860 $\mu$m data of
\citet{s08} while suffering less  from the effects of foreground
extinction than the 24 $\mu$m data of
\citet{so99}.
We use these new data  to derive more
precise measurements of the dust temperature, source size, and  luminosity
for each nucleus; although derived values depend heavily on the poorly
resolved geometry of the source, 
the luminosity surface densities also point to the
possible presence of an AGN in the western nucleus.

\section{Observations}

\begin{figure}
\includegraphics[scale=0.6]{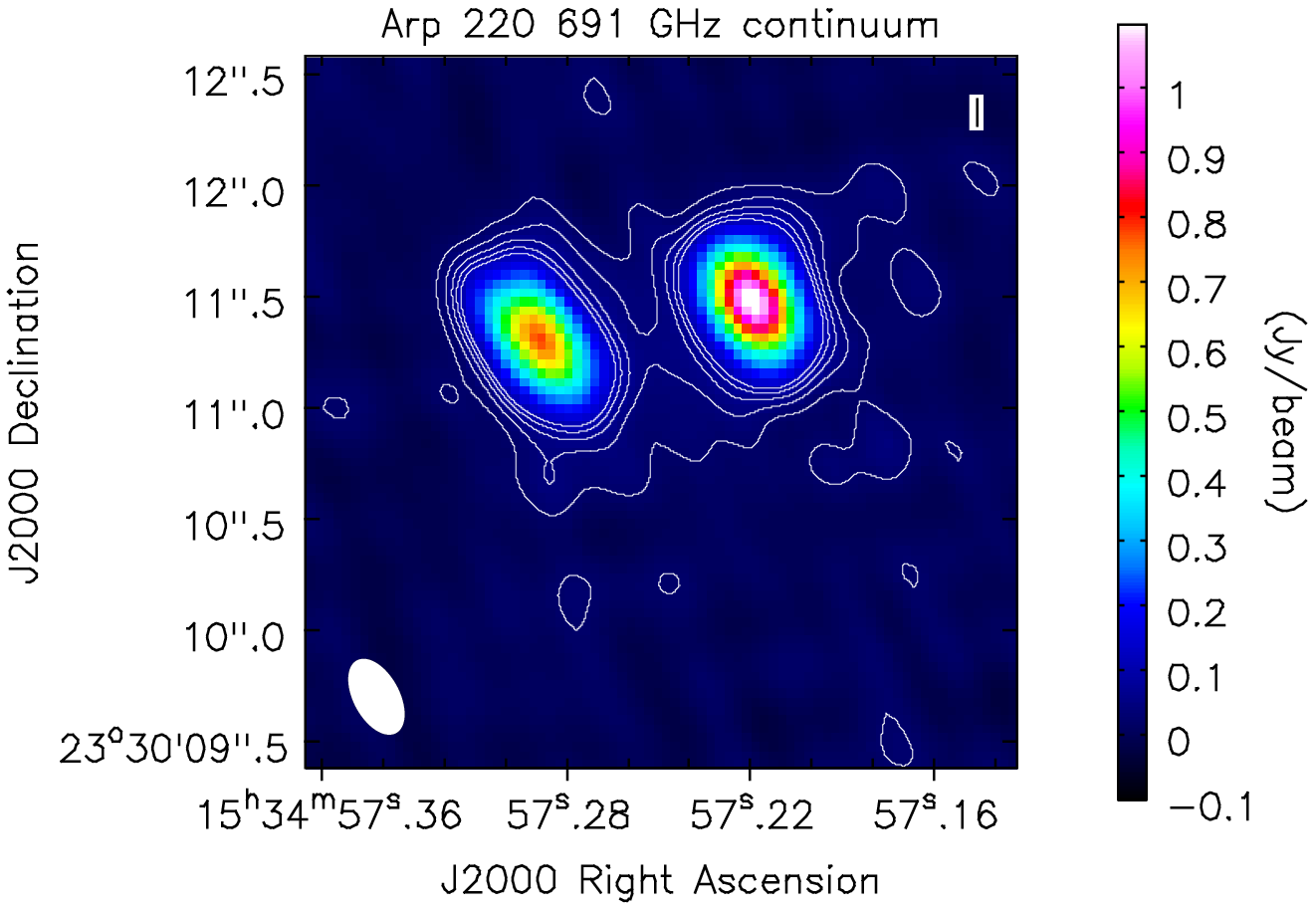}
\caption{The 691 GHz continuum image of Arp 220 obtained with
  ALMA with a resolution of $0.363\times 0.199^{\prime\prime}$ and
$1\sigma$ sensitivity 11 mJy beam$^{-1}$. 
Contours at
  $(2,4,6,8,10)\sigma$ show the extended low-level emission; the peak in
  the western nucleus is 1.15 Jy beam$^{-1}$.
The two nuclei appear
  as compact continuum sources and dominate the total flux; however,
  low-level emission on scales of 1-2$^{\prime\prime}$ contributes
  up to 24\% of the total flux  in this map.\label{image}}
\end{figure}

Arp 220 was observed with ALMA on 2012 December 31 with the Band 9
receivers tuned to cover the CO J=6-5 transition 
and adjacent continuum. Further details of the observations and 
calibration are given in \citet{r14}.  The observations covered
projected baselines from 13 to 374 m.
Because the absolute flux scale had to be set using
observations of the bandpass calibrator  (3C279, measured to be 6.0 Jy
during another observation taken within one day), we estimate
the absolute 
calibration accuracy to be no better than 15\%. 
 We combined two line-free regions of the spectrum centered at 680.90
GHz
($\Delta\nu = 1.375$ GHz) and 676.75 GHz ($\Delta\nu = 1.6875$ GHz)
to make the continuum image. 
The mean observed frequency of the resulting image is 678.755 GHz,
which at the redshift of Arp 220 (0.01813) corresponds to a rest
frequency of 691.061 GHz (434 $\mu$m). We
applied two rounds of phase-only self-calibration followed by a final
round of phase and amplitude  self-calibration. The final image (Figure
~\ref{image}) made
using uniform weighting and $0.04^{\prime\prime}$ pixels has a beam of 
$0.363\times 0.199^{\prime\prime}$ at position angle 28.3$^o$ with
a 1$\sigma$ noise level of 11 mJy beam$^{-1}$ (0.41 K). 
The two emission peaks are detected with a peak
signal-to-noise ratio of 100 in the western nucleus and 70 in the
eastern nucleus.
Observed and derived properties of the two nuclei are given in
Table~\ref{table1}. 

\section{Optical depth, source size, and dust temperature}

A $1\times 1^{\prime\prime}$ region centered on the peak pixel of each
nucleus was fit with an elliptical 
gaussian to obtain the observed size and total flux density. 
The fitted size 
was analytically deconvolved from the beam to obtain the true
source size.  For our primary analysis, we assume each nucleus has an
elliptical gaussian flux distribution. 
We also present selected results obtained by
adopting a model of a uniform elliptical disk, for which the
disk diameter is equal to 1.60 times the deconvolved gaussian
full-width half-maximum (FWHM) diameter \citep{s08}.  For the eastern nucleus,
the major axis of the deconvolved dust emission is 
aligned with the kinematic major axis of the CO emission
\citep{s08,r14}. The western nucleus is not well resolved along the
beam major axis, with a deconvolved diameter just 55\% of the beam;
in addition, we have no way to correct for any effects of radio seeing 
\citep[e.g. ][]{s08}. 
Thus, for the western nucleus we assume a major axis of
$0.21^{\prime\prime}$ at a position angle of 118$^\circ$,
which is consistent with the
kinematic major axis in this source. For our analysis, we also 
adopt the (deconvolved) inclination angle of
$53.5^\circ \pm 0.1^\circ$ obtained by \citet{b14} from a
two-dimensional non-linear least-squares fit of a thin, tilted,
exponential disk to their
33 GHz radio continuum image  with $0.08\times 0.06^{\prime\prime}$ resolution.
Our assumed orientation for the western nucleus agrees with the
east-west distribution of compact radio supernovae and supernova
remnants \citep{p07}.

The observed Rayleigh-Jeans peak temperature, $T_{obs}$, was converted
to a true  peak Rayleigh-Jeans brightness
temperature, $T_{B,peak}$, by considering the coupling of  a
  gaussian source to the gaussian beam,
$T_{obs} = T_{B,peak} \theta_{source}^2/\theta_{obs}^2$
where $\theta_{source}$ and $\theta_{obs}$ are, respectively, the true
and observed full-width
half-maximum diameter of the source.
We derive a peak brightness temperature for the eastern nucleus of $T_{B,peak} =
53 \pm 8$ K, where the uncertainty is the 1$\sigma$ uncertainty
derived from the 15\% calibration accuracy of the ALMA data. Because
the western nucleus is poorly resolved in one direction, its
peak brightness temperature is more uncertain. Using the measured sizes
gives $T_{B,peak} \ge 117 \pm 18$ K, while assuming the inclination
angle from \citet{b14} gives $T_{B,peak} = 181 \pm 27$ K. We thus
adopt
$T_{B,peak} = 181 \pm 27 ^{+0}_{-55} $ K, where the first uncertainty
is the $1\sigma$ uncertainty due to calibration and the second is the
systematic uncertainty due to the unknown minor axis size.\footnote{In
comparison, if we model each source as a uniform elliptical disk, the
derived brightness temperatures are somewhat lower ($T_B = 32$ K for the eastern nucleus and
$T_B =64$ K for the western nucleus using the observed source sizes).}

In general, deriving the true dust temperature, $T_D$, from the brightness 
temperature requires a knowledge of the optical depth, $\tau$, since
the two temperatures are related via
\begin{equation}
T_B = {h\nu/k \over e^{h\nu/kT_D} -1} (1- e^{-\tau})
\end{equation}
We estimate the optical depth at 691 GHz for each nucleus by comparing
our fluxes with previous measurements at 230 GHz from \citet{s99} and at
345 GHz from \citet{s08}. Because the sources are barely resolved, we
assume a constant optical depth for each source, which in turn implies
a constant mass surface density across each source.
We correct the 230 GHz fluxes for
contamination by synchrotron and free-free emission. For the western
nucleus, these fluxes are estimated to be 2.7 mJy and 10 mJy,
respectively \citep{de07}; for the eastern nucleus, we scale by the
relative 43 GHz fluxes of the two sources from \citet{rr05} to obtain
estimates of 2 and 9 mJy, respectively. We also correct the 345 GHz
fluxes for free-free emission by 10 and 9 mJy for the western and eastern
nucleus, respectively. We assume a dust emissivity
$\kappa_\nu = \kappa_o (\nu/\nu_o)^{1.8}$ \citep{p11} as seen towards
high column density lines of sight with the Planck satellite. We
measure the 691 GHz continuum flux for each nucleus within  a radius
of 0.45$^{\prime\prime}$ using an image with
0.5$^{\prime\prime}$ resolution to match the data from \citet{s08}.
We determine the optical depth using the 691 and 345 GHz fluxes from
\begin{equation}
 {F_{\nu_1} \over F_{\nu_2}} = \left( {\nu_1 \over \nu_2}\right)^3 \left( {e^{h\nu_2/kT_D}
-1 \over e^{h\nu_1/kT_D} -1}\right)
\left( {1 - e^{-\tau_1} \over 1 - e^{-\tau_2}}\right)
\end{equation}
where $F_\nu$ is the measured flux at frequency $\nu$ and $\tau_1 =
(\nu_1/\nu_2)^\beta \tau_2$. We 
use the lower signal-to-noise 230 GHz measurements as a
consistency check on the fits.
The solutions give a high 691 GHz optical depth in the western
nucleus and thus the solutions for $\tau$ and $T_D$ are decoupled. For
the eastern nucleus, we iterated between the two equations to find
consistent solutions for the two parameters  
(Table~\ref{table1}). \footnote{For the uniform elliptical disk model, the
dust temperatures are somewhat lower (55 K for the eastern nucleus
and 79 K for the western nucleus).}


The optical depths for the two nuclei (Table~\ref{table1}) are larger
than the value of $\tau_{250} = 2.7 \tau_{434} = 1$
derived from a global spectral energy distribution \citep{r11}.
 The dust temperature for the eastern nucleus agrees within
uncertainties with the global average value of 67 K \citep{r11}, while
the temperature for the western nucleus is roughly a factor of 2-3 higher.
The  691
GHz flux attributed to the two nuclei is $3.32 \pm 0.50$ Jy (Table~\ref{table1})
while the total flux 
measured from the ALMA map is $4.37\pm 0.66$ Jy. In comparison, the extrapolated flux
from the {\it Herschel} data is $5.9 \pm 0.9$ Jy
\citep{r11}. Thus, there is room for an extended, low surface
brightness dust component that could dominate the optical depth
determined from the
global fit. Indeed, the CO emission is significantly more extended
than the continuum emission \citep{s99,de07,s08,k12,r14} and there would
presumably be dust mixed in with this more extended gas. 

\section{Discussion}

\subsection{Revisiting the evidence for a black hole in the western nucleus of Arp 220}

Using our source sizes and dust temperatures, we can estimate the
luminosity, luminosity-to-mass ratio, and 
luminosity surface density for each of the two nuclei. 
 Integrating over both sides of the disk, 
the total luminosity of a thin inclined disk with a gaussian temperature
distribution is $L = (\pi \theta_{maj}^2)/8 \ln 2) \sigma
T_{D,peak}^4$ where $\theta_{maj}$ is the source full-width
half-maximum size along the major axis.
Depending on its minor axis size, the western nucleus may be up to
an order of magnitude more luminous than the eastern nucleus
(Table~\ref{table1}) and the resulting total luminosity is 
$\log L_{nuclei} = 11.87^{+0.25,0}_{-0.27,-0.61}$, where the first
uncertainty is the $1\sigma$ calibration uncertainty and the second is
the systematic uncertainty from the unknown minor axis of the western nuclei.
We also adopt a disk geometry in calculating the total luminosity
of Arp 220
\citep[see, e.g., ][]{s97}, giving a global luminosity $\log L = 11.95
\pm 0.03$ 
\citep[a factor of two smaller than the luminosity calculated assuming
spherical geometry by ][]{r11}.
The two nuclei together 
account for $83^{+65,+0}_{-38,-34}$\%  of the total luminosity of Arp
220. Thus, within our calibration and systematic uncertainties, the
two nuclei in Arp 220 likely contribute at least 45\% 
and may contribute nearly all of the total infrared flux.

The flux or luminosity surface density, $\sigma T^4$, is a key indicator for
distinguishing 
between starbursts and AGN. We note that there are inconsistencies in
the literature concerning this quantity, which is sometimes calculated
by combining a spherical luminosity estimate with a projected source
area, resulting in values a factor of 4 larger than the definition
used here.
 In the western nucleus of Arp 220, the 
luminosity surface density is  $2.1 \times 10^{14}$ 
L$_\odot$ kpc$^{-2}$,
while in the eastern nucleus it is $5.8 \times 10^{12}$ 
L$_\odot$ kpc$^{-2}$ (see Table~\ref{table1} for uncertainties in
these quantities).
Converting to the flux definition used here, the peak temperature of
162 K from 
\citet{s08} corresponds to  a luminosity surface density of
$1\times 10^{14}$ L$_\odot$ kpc$^{-2}$ for the western nucleus,
quite similar to the value obtained here value.
While  \citet{de07} quote a
flux of  $5\times 10^{14}$ L$_\odot$ kpc$^{-2}$ 
for the western nucleus,  this value combines a spherical estimate of
the luminosity with the projected area of the disk on the sky.
Using our
definition of flux and the peak dust
temperature of 90 K from a gaussian source model, the \citet{de07} data
give a flux of $1\times 
10^{13}$ L$_\odot$ kpc$^{-2}$. However, we note that at 1.3 mm  (and
possibly at 850 $\mu$m) the true dust 
temperatures and therefore the fluxes are almost certainly
underestimated due to optical depth effects.


The luminosity-to-mass ratio is another key indicator used to
distinguish between starbursts and AGN.
\citet{e11} model the CO emission as a thin rotating
disk to obtain  masses of $1.6 \times 10^9$ M$_\odot$
for the western nucleus within 100 pc (with 80\% of the mass within 50
pc radius) and $1.8 \times 10^9$
M$_\odot$ for the eastern nucleus within a radius of 81 pc.
Combining these masses with our derived luminosities
gives a luminosity-to-mass ratio of  540 L$_\odot$ M$_\odot^{-1}$ within
$r=50$ pc for the
western nucleus and 30 L$_\odot$ M$_\odot^{-1}$ within $r=80$ pc for 
the eastern nucleus (see Table~\ref{table1} for uncertainties). 
Our luminosity-to-mass ratio for the western nucleus is comparable to
that of \citet{s08}, who  obtained  a ratio $\ge 400$ L$_\odot$
M$_\odot^{-1}$ within $r = 40$ pc  using a
dynamical mass of $5.4 \times 10^8$ M$_\odot$ and a luminosity derived
assuming a spherical geometry.

The observed luminosities could be provided by a 
black hole with a mass of a few $\times 10^6$ M$_\odot$
accreting at the Eddington limit of $4\times 10^4$ L$_\odot$ M$_\odot^{-1}$. 
\citet{e11} use the $M_{BH}-\sigma$ relation to
estimate a black hole mass of $1.4 \times 10^8$ M$_\odot$ in the
western nucleus. We note that \citet{b11}
identify two compact radio sources that could be associated with a jet
from an AGN in the western nucleus. 
In the eastern nucleus, an extreme starburst, with
a maximum luminosity-to-mass ratio of $\sim 1000$ L$_\odot$
M$_\odot^{-1}$ and luminosity surface density of $10^{13}$ L$_\odot$
kpc$^{-2}$ \citep{t05,s08}, 
could also account for our observed values. 
While the luminosity surface density of the western nucleus excludes a
starburst solution at the $\sim 3\sigma$ level, the
derived values depend heavily on the poorly resolved geometry of the
source. We also note that a more extreme ``hot'' starburst
can produce even higher luminosity surface densities \citep{at11};
with a peak temperature of $\sim 200$ K, the western nucleus could be
probing this extreme starburst regime.

\subsection{Gas and star formation in the Arp 220 nuclei}


We made a first estimate of the gas surface density, $N_H$,
 perpendicular to the plane of the disk from the
optical depth using the value of $\tau_{250\mu m}/N_H = 2.32
\times 10^{-25}$ cm$^2$ derived
from Planck data \citep{p11} and using $\beta=1.8$ to derive
$\tau_{434\mu m}/N_H$. However, this surface density
results in a gas mass  for the western nucleus
that is  twice as large as the
dynamical mass within the same radius \citep{e11}.
These results suggest that
the dust optical depth per hydrogen atom, $\tau/N_H$, must be larger
in Arp 220 than the 
value from \citet{p11}, which was measured
towards the Taurus molecular cloud.
We note that the dust emissivity $\kappa_o$ is found to increase
at higher densities in the dust grain models of
\citet{oh94}, although the models do not probe the combination of high
density and temperature which we find in Arp 220 (see below).
 A smaller gas-to-dust mass ratio in Arp 220 than in the Milky Way
could also produce a larger value of $\tau/N_H$.
Adopting $\tau_{250\mu m}/N_H = 2 (\tau_{250\mu m}/N_H)_{Planck} =  4.6
\times 10^{-25}$ cm$^2$, we obtain the (uniform) gas surface densities 
given in Table~\ref{table1}. 
In particular, the gas surface density of
$N_H = 1.8 \times 10^{25}$ cm$^{-2}$ in the western nucleus implies
that any buried AGN would be Compton-thick 
\citep[see also ][]{s99,de07,s08}.

 Since the mean volume density of a dust disk depends on its vertical size
(which is unconstrained from our data), we use a spherical source
model to place lower limits on the volume density 
of $\ge 9\times 10^4$ cm$^{-3}$ for the
western nucleus and $\ge 2\times 10^4$ cm$^{-3}$ for the eastern nucleus.
These mean densities are significantly
higher than the density of either the cold or the warm gas
component traced via the CO analysis presented in \citet{r11}. 
Some of this difference
may be attributed to the more extended nature of the warm CO
emission seen in the CO 6-5 map \citep{r14}. However, the continuum
maps show that a significant fraction of the gas mass in Arp 220
occurs in a very centrally concentrated, high density component that
is not traced by the high-J CO lines.

These new observations of Arp 220 reveal many
similarities in the dust properties of the two nuclei. Both contain a
compact, high surface density disk with 
a high mean gas density and warm dust temperatures. 
While the data provide tantalizing clues for the presence of an 
AGN in the western nucleus, 
higher resolution observations with ALMA using
baselines of a few kilometers could resolve a 
temperature gradient in the central disk and reveal the high
luminosity surface density that is a characteristic signature of an
AGN.

\acknowledgments

We thank the referee for a thorough review which improved this paper
significantly. 
This paper makes use of the following ALMA data:
ADS/JAO.ALMA\*2011.0.00403.S. ALMA is a partnership of ESO
(representing its member states), NSF (USA) and NINS (Japan), together
with NRC (Canada) and NSC and ASIAA (Taiwan), in cooperation with the
Republic of Chile. The Joint ALMA Observatory is operated by ESO,
AUI/NRAO and NAOJ. The research of C.D.W. is supported by grants from
the Natural Sciences and Engineering Research Council of
Canada. C.D.W. also thanks the European Southern Observatory and the
National Radio Astronomy Observatory for visitor support.



{\it Facilities:} \facility{ALMA}.




\clearpage



\clearpage







\clearpage

\begin{deluxetable}{l|l|l|l}
\tablecaption{ALMA Arp 220 691 Gz (434 $\mu$m) Continuum Data\label{table1}}
\tablehead{\colhead{Quantity}&  \colhead{Eastern
    Nucleus} & \colhead{Western Nucleus}& units}
\startdata
Position (RA, Dec) & 
(15:34:57.29, +23:30:11.3) &   (15:34:57.22, +23:30:11.5) & J2000 \\ 
Peak temperature\tablenotemark{a,b} & 27.0 & 40.1 & K \\
Flux density (691 GHz)\tablenotemark{b} & 1.51 & 1.81 & Jy \\
Observed size\tablenotemark{b} & $ 0.50 \times 0.30$, 36$^o$ & $0.41 \times 0.29$, 28$^o$ & $\prime\prime$ \\
Flux density (691 GHz)\tablenotemark{c} & 1.45 & 1.82  & Jy \\
Flux density (345 GHz)\tablenotemark{d} & 191 & 370 & mJy \\
Flux density (230 GHz)\tablenotemark{e} & 55& 129 & mJy \\
\hline
Deconvolved size\tablenotemark{f} & 0.34 $\times$
0.22, 33$^o$ &  $0.21 \times \le 0.19$, 118$^o$  & $\prime\prime$ \\
 $\tau_{434\mu m}$\tablenotemark{g,h} & $1.7^{+1.1}_{-0.7}$ & $5.3^{+9.0}_{-1.7}$ & ... \\
$T_{D,peak}$\tablenotemark{h} & $80^{+30}_{-17} $ & $197 \pm 27 ^{+0}_{-55}$ & K \\
\hline
log(Luminosity)\tablenotemark{h} & $10.7^{+0.5}_{-0.4}$ & $11.8 \pm
0.2 ^{+0}_{-0.7}$ & L$_\odot$ \\
log(Flux = $\sigma T^4_{D,peak}$) & $12.8^{+0.5}_{-0.4}$
& $ 14.3\pm 0.2 ^{+0}_{-0.7}$ & L$_\odot$ kpc$^{-2}$ \\
$L/M_{dyn}$\tablenotemark{i} & $30^{+70}_{-20}$ &
$540^{+360,+0}_{-240,-240} $ & L$_\odot$  M$_\odot^{-1}$ \\ 
 Gas Mass\tablenotemark{i,j} & $1.1 \times 10^9$ & $1.3\times 10^9$ & M$_\odot$ \\
 Gas Surface Density\tablenotemark{j} & $6.4
 \times 10^{24}$ & $1.8 \times 10^{25}$ & H cm$^{-2}$ \\
 Gas Surface Density\tablenotemark{j} & $5.4
 \times 10^4$ & $1.4 \times 10^5$ & M$_\odot$ pc$^{-2}$ \\
 Mean Volume Density\tablenotemark{i,j,k} &
 $\ge 2\times 10^4$ & $\ge 9\times 10^4$ & H cm$^{-3}$ \\

\enddata

 {Note: calibration uncertainties for all fluxes are
  estimated to be 15\%.}
\tablenotetext{a}{Rayleigh-Jeans temperature. For this resolution and
  frequency, 1 Jy = 36.6 K.} 
\tablenotetext{b}{From fitting each source separately with an 
  elliptical gaussian;  691 GHz is the rest frequency.
  Reported size is FWHM; synthesized beam is $0.363^{\prime\prime}\times
  0.199^{\prime\prime}$ at position angle 28$^o$. }
\tablenotetext{c}{Flux density (rest frequency 691 GHz) measured within
  0.45$^{\prime\prime}$ radius from a 
  0.5$^{\prime\prime}$ resolution image.}
\tablenotetext{d}{Flux density (rest frequency 345 GHz) from \citet{s08}
measured within
  0.45$^{\prime\prime}$ radius from a 
  0.5$^{\prime\prime}$ resolution image. Flux is corrected for free-free emission
  of 9 and 10 mJy in the eastern and western nuclei;
  see text.}
\tablenotetext{e}{Flux density (rest frequency 230 GHz)
  from \citet{s99}, corrected for free-free and
  synchrotron emission (eastern: 9 and 2 mJy; western: 10 and 2.7
  mJy);  see text.} 
\tablenotetext{f}{Value reported is FWHM for a deconvolved gaussian
  source; $1^{\prime\prime} = 361$ pc.}
\tablenotetext{g}{Calculated assuming $\beta = 1.8$; see text.
 Uncertainties are derived from calibration uncertainties.
Note that in the solutions,
  higher values of $T_D$ correspond to lower
  values of $\tau$.}
\tablenotetext{h}{Assuming   a gaussian brightness distribution 
and an  inclination for the western nucleus of 53.5$^\circ$;
  see text.}
\tablenotetext{i}{Calculated  for $r \le 50$ pc for the western nucleus
 and $r \le 80$ pc for the eastern nucleus.}
\tablenotetext{j}{Assumes $\tau/N_H = 2 (\tau/N_H)_{Planck}$;
  see text.}
\tablenotetext{k}{ Upper limit is calculated assuming a spherical geometry
(which
  is quite unlikely, especially for the eastern nucleus).}
\end{deluxetable}


\begin{thebibliography}{}
\bibitem[Aalto et al.(2009)]{a09} Aalto, S., Wilner, D., Spaans, M. et
  al. 2009, \aap, 493, 481
\bibitem[Andrews \& Thompson(2011)]{at11} Andrews, B. H. \& Thompson,
  T. A., 2011, \apj, 727, 97
\bibitem[Armus et al.(2007)]{a07} Armus, L., Charmandaris, V.,
  Bernard-Salas, J. et al. 2007, \apj, 656, 148
\bibitem[Baan \& Haschick(1995)]{bh95} Baan, W. A. \& Haschick,
  A. D. 1995, \apj, 454, 745
\bibitem[Barcos-Mu\~noz et al.(2014)]{b14}Barcos-Mu\~noz, L. et al.,
  2014, \apj, submitted 
\bibitem[Batejat et al.(2011)]{b11}Batejat, F., Conway, J. E., Hurley,
  R., et al. 2011, \apj, 740, 95
\bibitem[Clements et al.(2002)]{c02}Clements, D. L., McDowell, J. C.,
  Shaked, S., et al. 2002 , \apj, 581, 974
\bibitem[Contini et al.(2013)]{c13}Contini, M., 2013, \mnras, 429, 242
\bibitem[Downes \& Eckart(2007)]{de07} Downes, D. \& Eckart, A. 2007,
  \aap, 468, L57
\bibitem[Downes \& Solomon(1998)]{ds98} Downes, D. \& Solomon,
  P. M. 1998, \apj, 507, 615
\bibitem[Elmegreen et al.(2008)]{e08}Elmegreen, B. G., Klessen, R. S., \&
  Wilson, C. D. 2008, \apj, 681, 365
\bibitem[Engel et al.(2011)]{e11}Engel, H., Davies, R. I., Genzel, R.,
  et al. 2011, \apj, 729, 58
\bibitem[Genzel et al.(1998)]{g98} Genzel, R., Lutz, D., Sturm, E. et
  al. 1998, \apj, 498, 579
\bibitem[K\"onig  et al.(2012)]{k12} K\"onig  S., García-Mar\'in, M.,
Eckart, A., Downes, D., \& Scharw\"achter, J. 2012, \apj, 754, 58
\bibitem[Matsushita et al.(2009)]{m09} Matsushita, S., Iono, D.,
  Petitpas, G. R., et al. 2009, \apj, 693, 56
\bibitem[Nardini et al.(2010)]{n10} Nardini, E., Risaliti, G., Watabe,
  Y., Salvati, M., \& Sani, E., 2010 \mnras,  405, 2505
\bibitem[Ossenkopf \& Henning(1993)]{oh94} Ossenkopf, V. \& Henning,
  T. 1994, \aap, 291, 943
\bibitem[Parra et al.(2007)]{p07}Parra, R., Conway, J. E., Diamond,
  P. J. et al. 2007, \apj, 659, 314
\bibitem[Planck Collaboration(2011)]{p11} Planck Collaboration, 2011,
  \aap, 536, A25 
\bibitem[Rangwala et al.(2011)]{r11} Rangwala, N., Maloney, P. R.,
  Glenn, J. et al. 2011, \apj, 743, 94
\bibitem[Rangwala et al.(2014)]{r14} Rangwala, N., Wilson, C. D.,
  Glenn, J. et al. 2014, in prep
\bibitem[Rodr\'iguez-Rico et al.(2005)]{rr05}Rodr\'iguez-Rico, C. A., 
  Goss, W. M. , Viallefond, F. et al. 2005, ApJ, 633, 198
\bibitem[Sakamoto et al.(1999)]{s99} Sakamoto, K., Scoville, N. Z.,
  Yun, M. S. et al. 1999, \apj, 514, 68
\bibitem[Sakamoto et al.(2008)]{s08} Sakamoto, K., Wang, J., Wiedner,
  M., et al. 2008, \apj, 684, 957
\bibitem[Scoville et al.(1991)]{s91} Scoville, N. Z., Sargent, A. I.,
  Sanders, D. B., \& Soifer, B. T. 1991, \apj, 366, L5
\bibitem[Scoville et al.(1997)]{s97}Scoville, N. Z., Yun, M. S., \&
  Bryant, P. M., 1997, \apj, 484, 702
\bibitem[Scoville et al.(1998)]{s98} Scoville, N. Z., Evans, A. S.,
  Dinshaw, N. et al. 1998, \apj, 492, L107
\bibitem[Soifer et al.(1984)]{s84}Soifer, B. T., Neugebauer, G.,
  Helou, G., et al. 1984, ApJ, 283, L1
\bibitem[Soifer et al.(1999)]{so99}Soifer, B. T. et al., 1999, \apj,
  513, 207
\bibitem[Thompson et al.(2005)]{t05}Thompson, T. A., Quataert, E., \&
  Murray, N., 2005, \apj, 630, 167
\end{thebibliography}
\end{document}